# Embedding or bridging? AI4Science teams differ more in knowledge recombination than in problem novelty

Yun Fu[1], Jielan Ding[1,2*]

1. National Science Library, Chinese Academy of Sciences, Beijing 100190, China.

2. Department of Information Resources Management, School of Economics and Management, University of Chinese Academy of Sciences, Beijing 101408, China.

[*] Corresponding author: dingjielan@mail.las.ac.cn

## Abstract

AI for Science (AI4Science) is reshaping scientific research, yet how the organization of AI-related and domain-scientific work relates to different dimensions of research performance remains poorly understood. We analyze 489,409 AI4Science articles published between 1990 and 2025, identifying seven affiliation-based organizational configurations comprising three forms of embedding and four forms of cross-unit bridging. Focusing on 2006–2025, we examine two dimensions of research performance: knowledge creation, characterized by scientific problem novelty and knowledge-recombination breadth, and research impact, assessed through journal placement and citation impact. We find that organizational configurations differ only modestly in problem novelty but substantially in knowledge-recombination breadth. Cross-unit bridging is consistently associated with broader knowledge recombination, particularly in AI-oriented bridging, without a corresponding advantage in problem novelty. However, greater recombination breadth does not necessarily align with stronger research impact. Science-oriented and Multi-orientation bridging exhibit the most consistent positive associations with JCI Q1 journal placement and citation impact, although neither shows a statistically significant adjusted advantage in publication in the Elite-4 journals (Nature, Science, Cell, and PNAS). These research-impact differences remain largely intact after accounting for problem novelty and recombination breadth. Our findings reveal that organizational configurations are associated with distinct patterns of knowledge creation and research impact, rather than a single hierarchy of research performance. The results highlight the importance of considering organizational arrangements in relation to specific scientific objectives, rather than assuming that any one configuration is universally superior.

## 1. Introduction

Artificial intelligence (AI) has become an increasingly important component of scientific research, supporting activities ranging from data analysis and prediction to molecular design and experimental discovery (Wang et al., 2023). Its growing adoption has been associated with changes in scientific productivity, novelty, and impact, although these relationships vary across fields and research settings (Gao and Wang, 2024; Bianchini et al., 2025, 2026). Yet greater use of AI does not necessarily lead to broader scientific exploration. Recent evidence suggests that gains in individual scientific productivity and impact can coexist with a narrowing of collective research attention (Hao et al., 2026). While these studies have advanced our understanding of AI adoption and its consequences, they leave open an important organizational question: **how does the arrangement of AI-related and domain-scientific work relate to research performance, encompassing both knowledge creation and research impact?**

AI4Science research involves markedly different organizational arrangements. Computational work may be embedded within domain-scientific units, situated in AI-oriented units, integrated within explicitly hybrid units, or connected to scientific work through collaboration across organizational boundaries. These arrangements can be examined through the sub-institutional affiliations represented in scientific publications, including departments, laboratories, institutes, and research centers. We distinguish seven article-level organizational configurations, spanning organizational embedding and cross-unit bridging, according to the orientations of participating units and their organizational connections.

Such organizational heterogeneity is substantial and has evolved over time. Among 489,409 classifiable AI4Science articles published between 1990 and 2025, Science-unit embedding accounts for 62.1%, while cross-unit bridging accounts for 30.4%. Multi-orientation bridging represents more than half of bridging articles, and the relative prevalence of different configurations has changed considerably over time (Figure 1). These patterns show that AI4Science has developed through multiple organizational arrangements rather than a single dominant model. Whether these arrangements are associated with systematic differences in knowledge creation and research impact, however, remains unclear.

Research on scientific collaboration suggests that organizational arrangements may be associated with different aspects of knowledge creation. As knowledge becomes increasingly specialized, scientific research often requires combining complementary expertise distributed across individuals, disciplines, and organizational settings (Wuchty et al., 2007; Jones, 2009). Connections between computational and domain-scientific units may broaden the knowledge and methodological resources available for solving scientific problems (Hong and Page, 2004; Jeppesen and Lakhani, 2010). Such connections may also expose researchers to unfamiliar perspectives and alternative research directions, creating opportunities for scientific exploration (Burt, 2004; Fleming, 2001; Uzzi et al., 2013). These possibilities are related but distinct. A relatively established problem may be addressed through the recombination of distant bodies of knowledge, while a novel problem may draw on a comparatively concentrated knowledge base. Organizational configurations may therefore differ in their associations with problem novelty and knowledge-recombination breadth, without necessarily affecting both in the same way. This motivates our first research question:

**RQ1. How is organizational configuration associated with knowledge creation, as measured by scientific problem novelty and knowledge-recombination breadth?**

Organizational configurations may also differ in their associations with research impact. Access to complementary expertise can expand scientific opportunities, but collaboration across disciplinary and organizational boundaries also introduces demands for coordination, communication, and knowledge integration (Cummings and Kiesler, 2007; Leahey et al., 2017). Previous research suggests that team diversity can relate differently to scientific novelty and impact (Lee et al., 2015), while the organization of collaboration may shape whether team members jointly engage in conceptual rather than primarily technical tasks (Lin et al., 2023). Organizational arrangements associated with broader knowledge recombination or greater problem novelty may therefore not exhibit stronger research impact. Their associations may also differ across measures of journal placement and citation impact. We therefore ask:

**RQ2. How are the seven organizational configurations associated with research impact, as measured by journal placement and citation impact?**

Examining knowledge creation and research impact separately, however, leaves open how their observed differences are related. If configuration-related differences in research impact largely reflect variation in problem novelty or knowledge-recombination breadth, accounting for these measures should reduce the estimated impact contrasts between configurations. Conversely, the persistence of these contrasts would suggest that the two knowledge-creation measures provide an incomplete statistical account of configuration-related differences in research impact. This leads to our third research question:

**RQ3. To what extent do scientific problem novelty and knowledge-recombination breadth statistically account for configuration-related differences in research impact?**

To address these questions, we link affiliation-based organizational configurations to article-level measures of knowledge creation and research impact in a large-scale corpus of AI4Science publications. We first examine how the seven configurations differ in scientific problem novelty and knowledge-recombination breadth, then compare their associations with journal placement and citation impact. Finally, we assess whether configuration-related differences in research impact persist after accounting for the two knowledge-creation measures. The following sections develop the theoretical framework, describe the data and empirical strategy, and present the findings and their implications for science policy and research management.

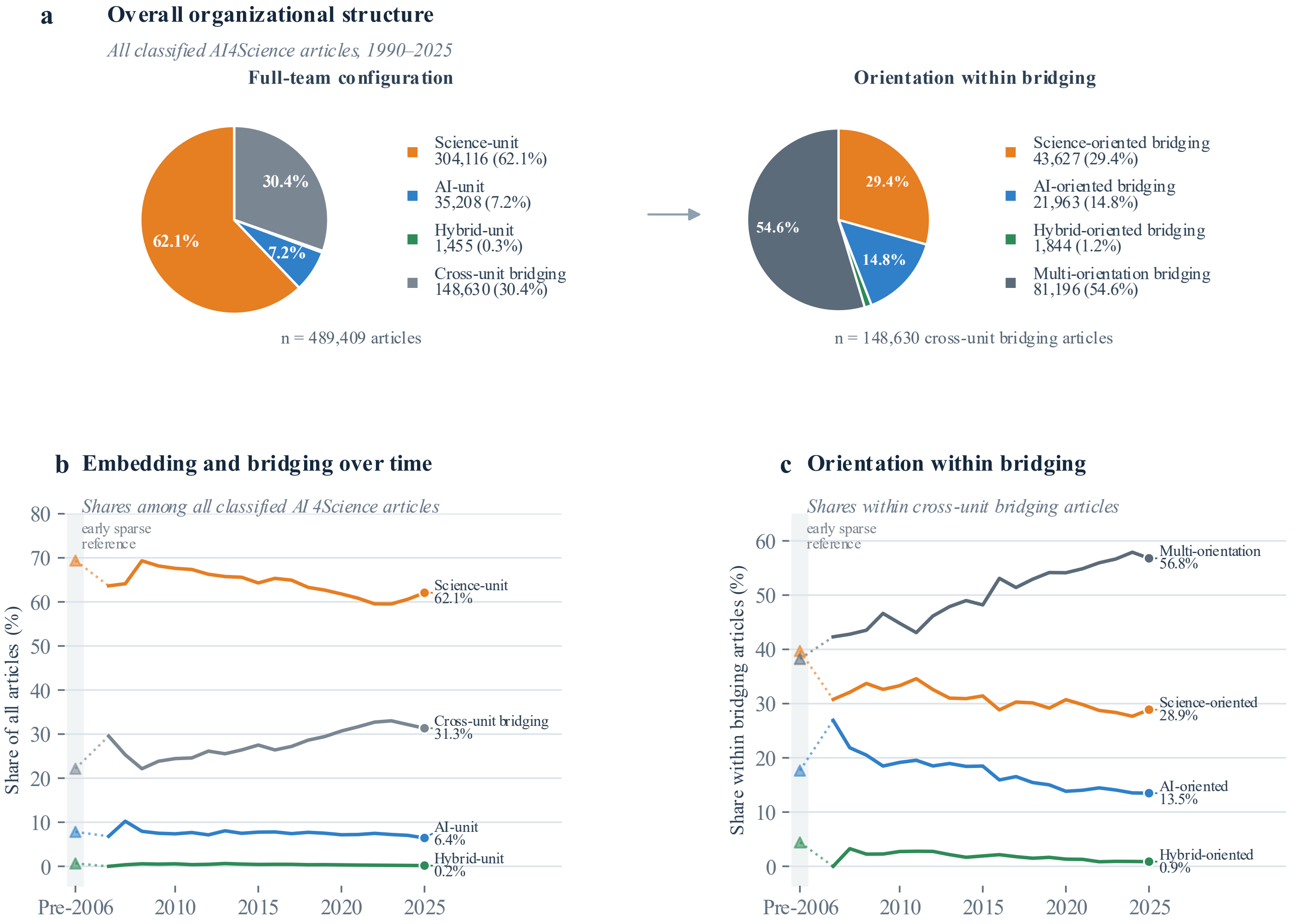

Figure 1. Evolution of organizational configurations in AI4Science. a, Overall organizational structure. b, Embedding and bridging over time. c, Bridging orientations over time.

## 2. Theoretical background and conceptual framework

Research on scientific search and organizational knowledge integration offers a basis for understanding the organizational heterogeneity observed in AI4Science. Different organizational arrangements may shape both the knowledge available to researchers and the conditions under which that knowledge is used. Yet access to diverse expertise, the pursuit of novel problems, and the production of influential research are not necessarily aligned. These distinctions are important for examining how organizational configurations relate to different dimensions of research performance and how these relationships differ.

### 2.1 Scientific search and knowledge mobilization

Scientific research involves identifying problems worth pursuing and assembling the knowledge needed to investigate them. Although closely related, these activities draw on partly different processes. Research on organizational search has shown that existing knowledge and cognitive representations influence which opportunities researchers recognize and how they evaluate unfamiliar alternatives (Gavetti and Levinthal, 2000;

Kaplan, 2008). Search is also shaped by the scope and depth of prior knowledge explored, including the extent to which researchers move beyond familiar knowledge domains or draw on knowledge developed at different times (Katila and Ahuja, 2002; Nerkar, 2003). The problems researchers choose and the knowledge they mobilize to address them therefore represent distinguishable dimensions of scientific search.

Organizational boundaries may matter for both dimensions, but not necessarily in the same way. Connections between different knowledge communities can introduce unfamiliar perspectives and alternative problem representations, potentially changing which scientific questions researchers recognize as worth pursuing. Yet the circulation of knowledge across boundaries can also transform how an existing problem is approached without changing the problem itself. Hargadon and Sutton (1997), for example, showed how technological innovations can arise from transferring established solutions between otherwise disconnected settings. In AI4Science, connecting computational and domain-scientific units may similarly provide new ways of investigating established scientific questions, even when the underlying research agenda remains largely unchanged.

A related literature emphasizes the role of organizational connections in locating and mobilizing knowledge. Hansen (1999) found that weak interunit ties facilitate knowledge search while making the transfer of complex knowledge more difficult. Borgatti and Cross (2003) showed that information seeking depends on awareness of relevant expertise and the perceived costs of accessing it. Collaboration networks can further expand opportunities to acquire knowledge through direct and indirect connections, although the benefits of network access depend on the capacity to absorb and use what is obtained (Ahuja, 2000; Tsai, 2001). Research on organizational diversity in scientific collaboration likewise points to relationships between the composition of collaborative networks and scientific outcomes (Yoo et al., 2024).

These studies suggest that organizational connections may be reflected more directly in the range of knowledge mobilized for research than in the novelty of the problems selected. For AI4Science, this distinction is particularly relevant because computational capabilities can support the investigation of existing scientific questions while also creating opportunities for new directions of inquiry. Whether different organizational configurations are more closely associated with one dimension than the other remains an empirical question. Comparing scientific problem novelty with knowledge-recombination breadth therefore provides a way to examine where organizational differences are most evident in the knowledge produced by AI-enabled research.

### 2.2 Organizational boundaries, knowledge integration, and research impact

The ability to access diverse knowledge does not necessarily imply the ability to combine it productively. The knowledge-based view of organizations emphasizes the integration of specialized expertise as a central source of organizational capability (Grant, 1996), while absorptive-capacity theory highlights the importance of prior related knowledge for recognizing and applying external knowledge (Cohen and Levinthal, 1990). Organizations also develop shared routines and combinative capabilities that support the use of accumulated expertise (Kogut and Zander, 1992). Even within established organizational settings, however, knowledge transfer can be impeded by ambiguity, limited understanding, and difficulties in adapting knowledge to new circumstances (Szulanski, 1996).

Crossing organizational boundaries changes the conditions under which knowledge is accessed and integrated. Connections between units may extend the range of available expertise, but they also require participants to communicate across different practices, representations, and expectations. The benefits of such connections depend on the structure and strength of organizational ties, as well as the processes through which specialized contributions are coordinated (Tortoriello et al., 2012). Experimental research has shown that structured interaction can improve groups' capacity to coordinate and integrate knowledge (Okhuysen and Eisenhardt, 2002). Institutional arrangements may also shape the collaborative connections through which researchers gain access to different knowledge communities (Barley et al., 2025).

These considerations provide a basis for distinguishing organizational embedding from cross-unit bridging. Embedding may facilitate knowledge exchange through shared organizational settings and established interaction routines, whereas bridging may connect more differentiated knowledge environments. Hybrid-unit embedding represents a further possibility in which AI-related and domain-scientific orientations coexist within a common organizational unit. None of these arrangements, however, can be assumed to provide uniformly better conditions for scientific work. Research on knowledge diversity and innovation has shown that combining different knowledge domains can change the distribution of innovative outcomes rather than simply increasing their average level (Taylor and Greve, 2006). Studies of interdisciplinary collaboration similarly emphasize the importance of collaborative practices and arrangements in realizing the potential of specialized expertise (Vladova et al., 2025).

The implications for research performance are therefore uncertain. Broad search and unusual knowledge combinations can be associated with influential research (Schilling and Green, 2011), but gaining access to external knowledge is distinct from incorporating it into productive scientific work (Dahlander and Gann, 2010). The benefits of interdisciplinary knowledge integration and diffusion also vary across research fields (Xu et al., 2025). Moreover, organizational diversity can have different relationships with citation impact and scientific disruption, depending on other team characteristics (Yoo et al., 2024). Research on the practical integration of AI into domain science further identifies communication and collaboration challenges that complicate the use of computational expertise in scientific work (Thu et al., 2025).

Organizational configuration may consequently be associated with research impact in different ways, depending on the knowledge involved, the demands of collaboration, and the outcome considered. Greater organizational heterogeneity need not correspond to stronger journal placement or citation impact, nor should all bridging configurations be expected to exhibit similar impact patterns. Comparing the seven configurations across multiple research-impact measures allows us to examine these differences without assuming a universal advantage for either embedding or bridging.

### 2.3 Relating knowledge creation to research impact

The relationship between organizational configuration and research performance cannot be understood solely from the presence of heterogeneous organizational participation. Scientific problem novelty and knowledge-recombination breadth describe different aspects of the knowledge represented in a publication, and neither necessarily captures the processes through which diverse contributions are integrated into a scientific result. Research on interdisciplinarity suggests that the diversity of cited knowledge and the disciplinary composition of author teams can exhibit different associations with research impact (Gates et al., 2024). Evidence from climate-change research likewise shows that collaboration across organizational levels relates differently to the variety and balance of referenced disciplines (Qiu et al., 2024). Knowledge creation and research impact should therefore be examined as related but empirically distinct dimensions of research performance.

This distinction raises an important question about the performance differences observed across organizational configurations. If these differences are closely associated with variation in the problems pursued or the knowledge mobilized, accounting for problem novelty and recombination breadth should alter the estimated configuration–performance relationships. Conversely, if the differences remain largely unchanged, these two characteristics provide only a limited statistical account of the performance patterns associated with organizational configuration.

We examine this possibility by comparing baseline performance models with models that additionally account for problem novelty, recombination breadth, and both measures together. This approach identifies whether the measured characteristics of knowledge production alter the estimated organizational differences in research performance. It does not establish that either characteristic mediates a causal effect of organizational configuration. Research problems, knowledge sources, organizational participation, and performance may be jointly influenced by factors that publication records cannot directly reveal.

The resulting framework treats organizational configuration as an observable aspect of how AI-related and domain-scientific work is situated and connected in scientific research. By examining its associations with knowledge creation and research impact, and assessing how the former measures statistically account for differences in the latter, the analysis seeks to characterize the organizational patterns of AI4Science beyond the distribution of organizational forms alone.

## 3. Data and methods

This article-level observational study examines the associations between organizational configuration and two dimensions of research performance in AI4Science: knowledge creation and research impact. The empirical workflow comprises four components: constructing the AI4Science corpus, identifying organizational configurations from author affiliations, measuring scientific problem novelty and knowledge-recombination breadth, and estimating configuration-related differences in journal placement and citation impact. We further examine how these impact differences change after accounting for the two knowledge-creation measures (Figure 2).

### 3.1 Data sources and corpus construction

We constructed an AI4Science corpus from the Science Citation Index Expanded (SCI-Expanded) and Social Sciences Citation Index (SSCI) collections of Web of Science (WoS). The study covers AI-related research in scientific fields outside core computer science, using articles as the unit of analysis. Our corpus definition includes a broad range of AI-related research rather than restricting inclusion to publications in which AI constitutes the primary research method or contribution.

To capture emerging AI terminology, we selected OpenAlex for its coverage of preprints alongside published literature, rather than relying exclusively on commercial citation databases. We assembled a reference corpus of OpenAlex-indexed publications from 2020 onward in four AI-related fields: Artificial Intelligence, Machine Learning, Computer Vision, and Natural Language Processing. Candidate expressions were extracted from titles and abstracts using 2–5-gram patterns and assessed according to frequency, co-occurrence, and temporal-growth characteristics. LLM-assisted assessment and human review were then used to evaluate their relevance and consolidate related expressions. The resulting vocabulary comprised 2,551 AI-related terms (Supplementary Data 1).

We applied this vocabulary to WoS Topic fields and screened the retrieved records for broad-AI relevance using DeepSeek-V4-flash. The screening considered AI-related methods, models, tools, systems, datasets, and evaluation based on article titles and abstracts, without requiring AI to be the paper's primary contribution. We excluded records without the WoS Article designation or valid WoS Category information, as well as those assigned exclusively to seven designated core computer-science categories. Articles also assigned to other subject categories were retained. The final corpus contains 503,282 articles published between 1990 and 2025. Further details of terminology construction and relevance screening are provided in Appendix A.2.

The full period is used to describe the development of AI4Science, whereas the main inferential analyses focus on 2006–2025, when publication volumes and affiliation information provide a more suitable basis for organizational comparisons. We divide the analysis into three periods (2006–2016, 2017–2022, and 2023–2025), using 2006, 2017, and 2023 as reference points associated with the rise of deep learning, the introduction of Transformer architectures, and the rapid expansion of large language models, respectively. These divisions provide temporal reference points for comparison rather than defining distinct or mutually exclusive technological phases.

The corpus was linked to author affiliations, organizational units, field and country assignments, research-task representations, journal indicators, and citation data. Citation benchmarks were constructed from the broader SCI-Expanded and SSCI records by WoS Category, publication year, and document-type group. Under the strict organizational-completeness criteria described in Section 3.2, organizational configurations were identified for 489,409 articles across 1990–2025, including 489,102 articles in the main analytical period.

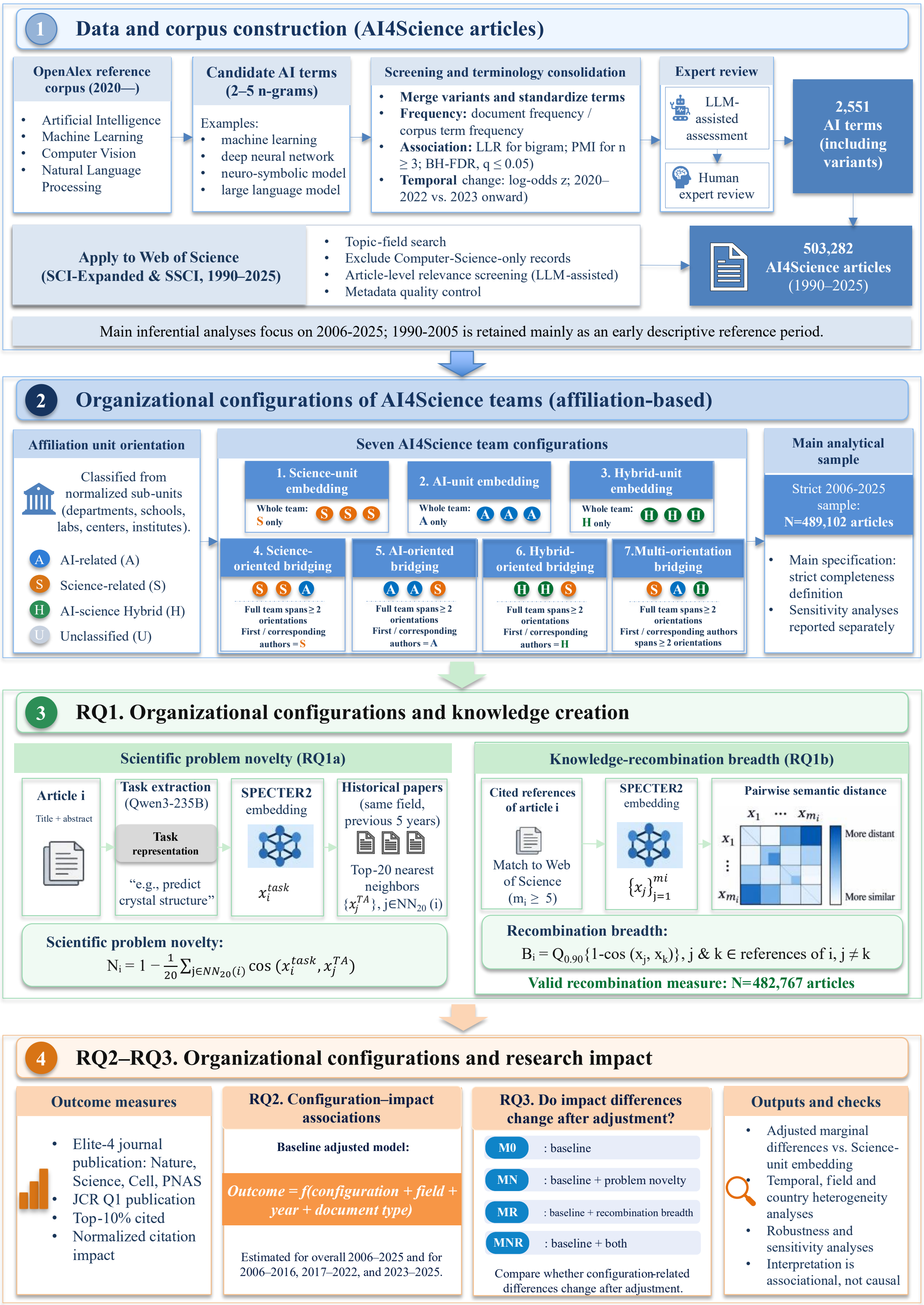

Figure 2. Empirical design and measurement workflow.

## 3.2 Organizational configuration

### 3.2.1 Affiliation-unit orientation

We identified organizational participation from the sub-institutional units reported in author affiliations, including departments, schools, laboratories, centers, and institutes. These units were classified at the lowest informative organizational level, rather than assigning orientations from the discipline or name of the parent institution. The classification therefore captures the organizational settings represented in a paper, not the disciplinary expertise of individual researchers or their specific contributions.

Each substantive unit was assigned one of three orientations: Science (S), AI (A), or Hybrid (H). Science refers to organizational units oriented toward substantive scientific domains, while AI refers to units explicitly oriented toward artificial intelligence or related computational methods. Hybrid is reserved for units whose own identity explicitly integrates AI or computing with a substantive non-computational scientific domain. The co-occurrence of AI and Science across different units, or merely across levels of an institutional hierarchy, is insufficient for a Hybrid classification. Hybrid is treated as a distinct orientation, rather than as simultaneous AI and Science labels.

Normalized affiliation-unit names and their organizational context were assessed through model-assisted classification, deterministic consistency checks, and additional review of ambiguous cases. Units lacking sufficient information for a substantive orientation were distinguished from strings that did not identify a substantive organizational unit. The former were treated as unresolved for the strict classification, rather than assigned an orientation by inference. A stratified review of 300 affiliation units confirmed 280 original classifications (93.3%), with most discrepancies involving units initially classified as Science (Appendix A.1; Supplementary Data 3).

### 3.2.2 Article-level organizational configurations

Organizational configuration was constructed from the orientations of the affiliation units represented in each article. For article $i$, let $U_i$ denote its substantive affiliation units and o(u) the orientation assigned to unit (u). The full-team orientation set is

$$O_i = \{o(u): u \in U_i\}. \qquad (1)$$

Articles with a single orientation ($|O_i| = 1$) are classified as embedding, while those with at least two orientations ($|O_i| \geq 2$) are classified as cross-orientation bridging. Embedding is classified as Science-unit, AI-unit, or Hybrid-unit when $O_i = \{S\}$, $\{A\}$, or $\{H\}$, respectively. For bridging articles, let $O_i^{FC}$ denote the orientations represented among first- and corresponding-author affiliations. Bridging is Science-oriented, AI-oriented, or Hybrid-oriented when $O_i^{FC} = \{S\}$, $\{A\}$, or $\{H\}$, respectively, and Multi-orientation bridging when $|O_i^{FC}| \geq 2$ (Table 1). The first-listed author was identified by position 1 in the WoS author list, and corresponding authors by the corresponding-author designation in WoS metadata.

Under the strict-completeness specification, all substantive affiliation units must have resolved orientations, including those associated with first and corresponding authors. The main analytical sample comprises 489,102 articles published between 2006 and 2025, including 1,453 Hybrid-unit embedding articles (0.30%). Estimates for this rare category warrant caution.

**Table 1. Operational definitions of the seven organizational configurations**

| Full-team orientations | First/corresponding-author orientations | Configuration |
|---|---|---|
| $O_i = \{S\}$ | — | Science-unit embedding |
| $O_i = \{A\}$ | — | AI-unit embedding |
| $O_i = \{H\}$ | — | Hybrid-unit embedding |
| $\lvert O_i \rvert \geq 2$ | $O_i^{FC} = \{S\}$ | Science-oriented bridging |
| $\lvert O_i \rvert \geq 2$ | $O_i^{FC} = \{A\}$ | AI-oriented bridging |
| $\lvert O_i \rvert \geq 2$ | $O_i^{FC} = \{H\}$ | Hybrid-oriented bridging |
| $\lvert O_i \rvert \geq 2$ | $\lvert O_i^{FC} \rvert \geq 2$ | Multi-orientation bridging |

### 3.3 Measures of knowledge creation

#### 3.3.1 Scientific problem novelty

Problem novelty measures how far the scientific problem addressed by an article departs from recent research in the same field, rather than directly assessing the novelty of its AI methods. We used Qwen3-235B to extract the principal research task from each article's title, abstract, and WoS-derived disciplinary information. The extracted tasks were subsequently used to measure scientific problem novelty. A review of 100 sampled articles found that 93% of the extracted tasks accurately captured the central research problem, while the remaining 7% retained the main research focus but were less precise in their descriptions (Supplementary Data 2). The extraction procedure, including the recovery of initially missing tasks, is detailed in Appendix A.3, with the complete prompts provided in Supplementary Note A.1.

Let $x_i^{task}$ denote the SPECTER2 embedding of article $i$'s extracted task. Its historical reference pool consists of WoS articles published in the preceding five years within the same JCR broad field(s), represented by SPECTER2 embeddings of their titles and abstracts, $(x_j^{TA})$. We identify the 20 most similar historical articles using cosine similarity and define problem novelty as

$$N_i = 1 - \frac{1}{20} \sum_{j \in NN_{20}(i)} \cos\left(x_i^{task}, x_j^{TA}\right). \qquad (2)$$

Higher scores indicate greater semantic distance from recent scientific precedents. Historical neighbors are retrieved through exhaustive search of the eligible reference pool. Sensitivity analyses examine alternative neighborhood sizes and task-based representations of historical papers (Appendix B.2).

#### 3.3.2 Knowledge-recombination breadth

Knowledge-recombination breadth captures the semantic dispersion of the prior literature cited by an article. Let $R_i$ denote its references matched to WoS records with SPECTER2 embeddings. For each distinct reference pair $j, k \in R_i, j < k$, cosine distance is calculated as

$$d_{jk} = 1 - \cos\left(x_j, x_k\right). \qquad (3)$$

Knowledge-recombination breadth is defined as the 90th percentile of these pairwise distances:

$$B_i = Q_{0.90}\left(\{d_{jk}: j, k \in R_i,\ j < k\}\right). \qquad (4)$$

Higher values indicate greater semantic dispersion among cited references, rather than their successful integration. At least five matched references are required, yielding 482,767 articles with valid scores during 2006–2025. Sensitivity analyses vary the distance percentile and minimum reference threshold (Appendix B.3).

### 3.4 Research-impact outcomes

We examine four article-level outcomes covering journal placement and citation impact. Journal placement is measured by Q1 status based on the Journal Citation Indicator (JCI) quartiles reported for the 2024 data year in Journal Citation Reports (JCR), and by publication in *Nature*, *Science*, *Cell*, or *Proceedings of the National Academy of Sciences* (Elite-4). Citation impact is measured by top-10% cited status and normalized citation score (NCS), using the broader WoS citation distributions by subject category, publication year, and document-type group as reference benchmarks. NCS is winsorized at the 99th percentile. These indicators capture different aspects of research impact; journal placement does not directly measure the quality of individual articles.

### 3.5 Empirical strategy

Across analyses, Science-unit embedding serves as the reference configuration for reported contrasts. Holm correction was applied separately within each outcome and analytical specification, covering all 21 pairwise configuration comparisons for knowledge-creation measures and six comparisons with the reference configuration for research-impact outcomes. Reported 95% confidence intervals are nominal.

#### 3.5.1 Organizational configuration and knowledge creation (RQ1)

For each knowledge-production outcome $Y_i \in \{N_i, B_i\}$, we estimate

$$Y_i = \alpha + \sum_{c \neq S} \beta_c \; Conf_{ic} + \lambda_{f(i) \times t(i)} + \varepsilon_i. \qquad (5)$$

Here, $Conf_{ic}$ indicates organizational configuration $c$, and $\lambda_{f(i) \times t(i)}$ denotes broad-field-by-publication-year fixed effects. OLS estimates use heteroskedasticity-robust standard errors and are reported in both original and sample-standard-deviation units. Analyses are conducted overall and by publication period, field, and country. Team size and collaboration scope are excluded from the primary specification because they are closely related to organizational configuration.

#### 3.5.2 Organizational configuration and research impact (RQ2)

Research-impact models include configuration, primary WoS Category, publication-year, and document-type fixed effects. Each outcome is estimated using all eligible articles with no missing data for that outcome. We use Firth bias-reduced logistic regression for the rare Elite-4 outcome, logistic regression for Q1 placement and top-decile citations, and OLS for winsorized NCS. Results are reported as adjusted probabilities and probability differences for binary outcomes (per 10,000 papers for Elite-4), and adjusted means and mean differences for NCS. Temporal, field, and country heterogeneity is examined through corresponding subgroup analyses.

#### 3.5.3 Accounting for problem novelty and recombination breadth (RQ3)

We compare four specifications on a common complete-case sample for each research-impact outcome: the baseline model ($M_0$), baseline plus problem novelty ($M_N$), baseline plus recombination breadth ($M_R$), and baseline plus both ($M_{NR}$). The signed change in the adjusted configuration contrast is

$$D(c, m) = \Delta_{c,m} - \Delta_{c,M_0}, \qquad m \in \{M_N, M_R, M_{NR}\}. \qquad (6)$$

Positive and negative values indicate upward and downward shifts, respectively; attenuation is identified by a reduction in the absolute contrast. These comparisons describe changes in adjusted associations, not causal mediation.

## 4. Results

### 4.1 Organizational configuration shows limited differentiation in scientific problem novelty

Despite the growing prevalence of cross-unit bridging across scientific fields and countries (Supplementary Fig. C1), organizational configuration shows only modest differentiation in scientific problem novelty. Across 2006–2025, adjusted differences relative to Science-unit embedding are small for all six alternative configurations (Figure 3a). AI-unit embedding has the largest positive difference (+0.030 SD), whereas the four bridging configurations cluster much closer to zero (−0.001 to +0.011 SD). The pooled results therefore provide little evidence that connecting different organizational orientations is consistently associated with the pursuit of more novel scientific problems.

This relationship has nevertheless changed over time. All six non-reference configurations move from negative point estimates in 2006–2016 to positive estimates in 2023–2025 (Figure 3b). The upward shift extends to both embedding and bridging configurations, rather than favoring a particular form of cross-orientation collaboration. Moreover, the recent differences remain small, suggesting a change in the relative positioning of organizational configurations rather than the emergence of a strong novelty advantage.

Scientific fields further qualify this pattern. Configuration-related differences vary considerably across disciplines, but positive average contrasts do not necessarily favor bridging over embedding (Figure 4). Agricultural Sciences, Plant & Animal Science, and Environment/Ecology, for example, experienced some of the largest increases in bridging prevalence, yet their field-level comparisons show negative mean bridging–embedding differences. Nor is a consistent positive bridging pattern evident across the ten leading contributing countries (Supplementary Fig. C2). Thus, neither the expansion of organizational bridging nor the variation in its disciplinary setting translates into a general advantage in scientific problem novelty.

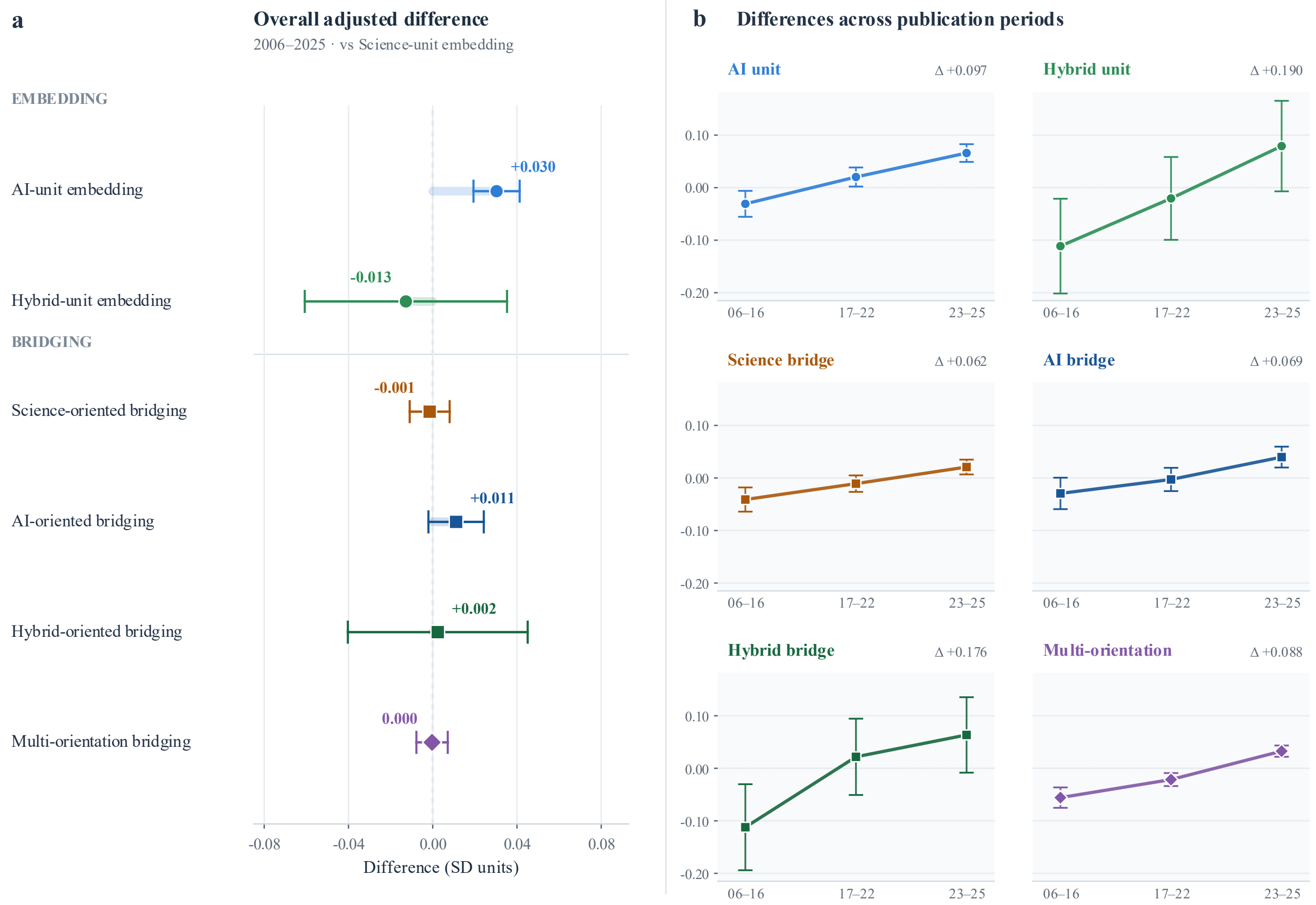

Figure 3. **Organizational configuration and scientific problem novelty.** a, Overall adjusted differences. b, Differences across publication periods.

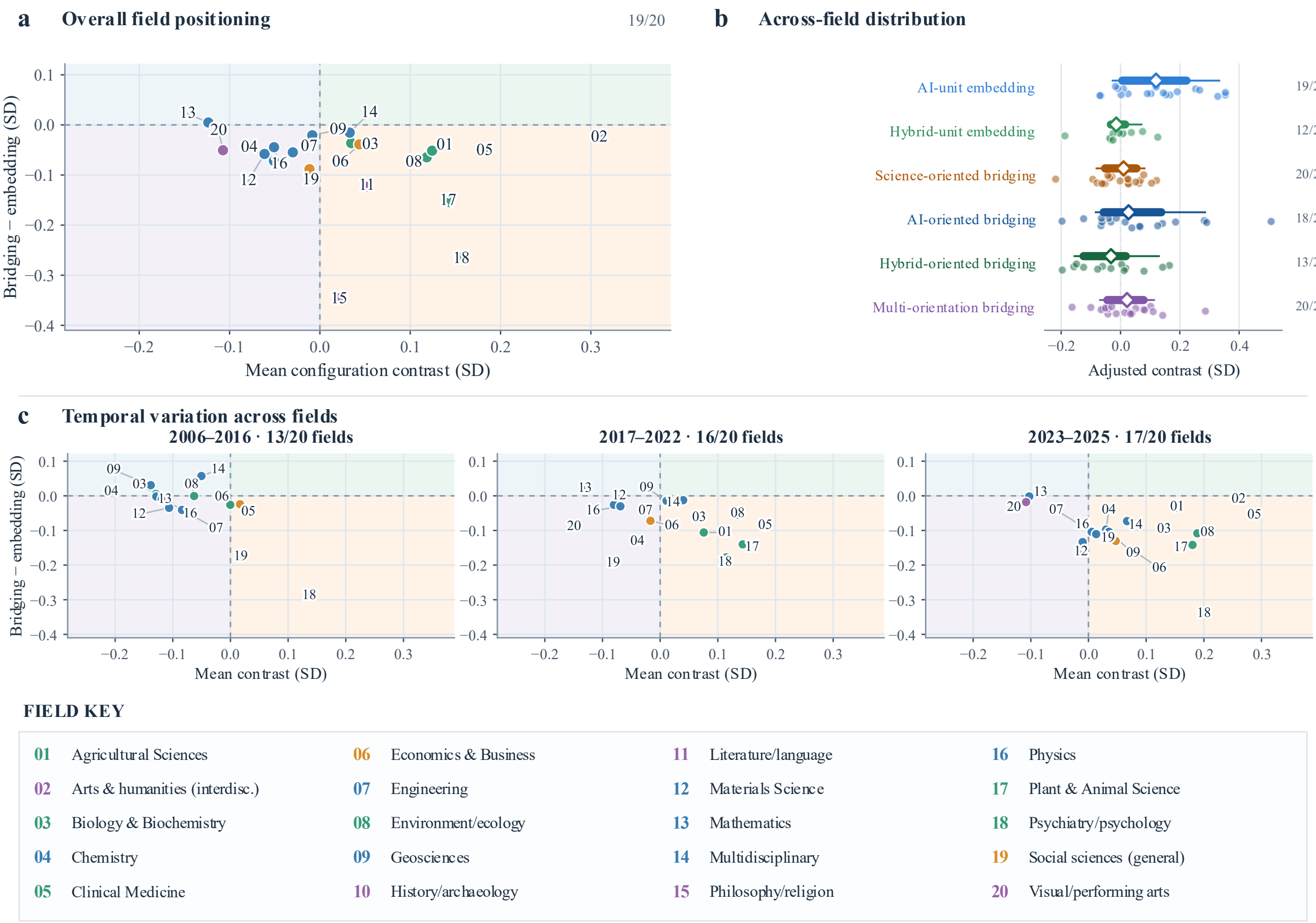

Figure 4. **Field heterogeneity in scientific problem novelty.** a, Overall field positioning. b, Across-field distributions. c, Temporal variation across fields.

### 4.2 Organizational configuration is more strongly associated with knowledge-recombination breadth

A substantially clearer organizational pattern emerges for knowledge-recombination breadth. Across 2006–2025, all six non-reference configurations show positive adjusted differences relative to Science-unit embedding (Figure 5a). The largest difference occurs in AI-oriented bridging (+0.197 SD), followed by Hybrid-oriented (+0.160 SD), Multi-orientation (+0.140 SD), and Science-oriented bridging (+0.115 SD). By comparison, AI-unit and Hybrid-unit embedding show smaller differences (+0.080 and +0.057 SD, respectively). Unlike problem novelty, recombination breadth therefore distinguishes bridging configurations more clearly from organizational embedding.

These differences also show a more persistent temporal pattern (Figure 5b). Several bridging configurations are already positively associated with recombination breadth in 2006–2016, with substantially larger differences emerging in 2017–2022. Although the estimates generally decline in 2023–2025, all four bridging configurations retain positive contrasts relative to Science-unit embedding. The temporal pattern is therefore not one of continuous growth, but of a recurring association between cross-orientation bridging and broader knowledge recombination.

Field- and country-level analyses reinforce this distinction while revealing variation in its strength. In contrast to problem novelty, mean bridging–embedding differences in recombination breadth are positive in many scientific fields (Figure 6a,c). Science-oriented and Multi-orientation bridging also exhibit positive field-level point estimates across all 20 fields (Figure 6b). The pattern extends across countries: in 2023–2025, Science-oriented, AI-oriented, and Multi-orientation bridging have positive point estimates in all ten leading contributing countries, although their precision varies (Supplementary Fig. C3). These findings suggest a broadly recurring association between bridging and recombination breadth, rather than an advantage confined to a small number of disciplinary or national settings.

The contrast between problem novelty and knowledge-recombination breadth is clear. On comparable standardized scales, organizational configurations differ more substantially in recombination breadth than in problem novelty, with larger and more consistently positive differences among bridging configurations. Organizational bridging is thus more closely associated with the breadth of prior knowledge mobilized in scientific work than with the novelty of the problems being pursued.

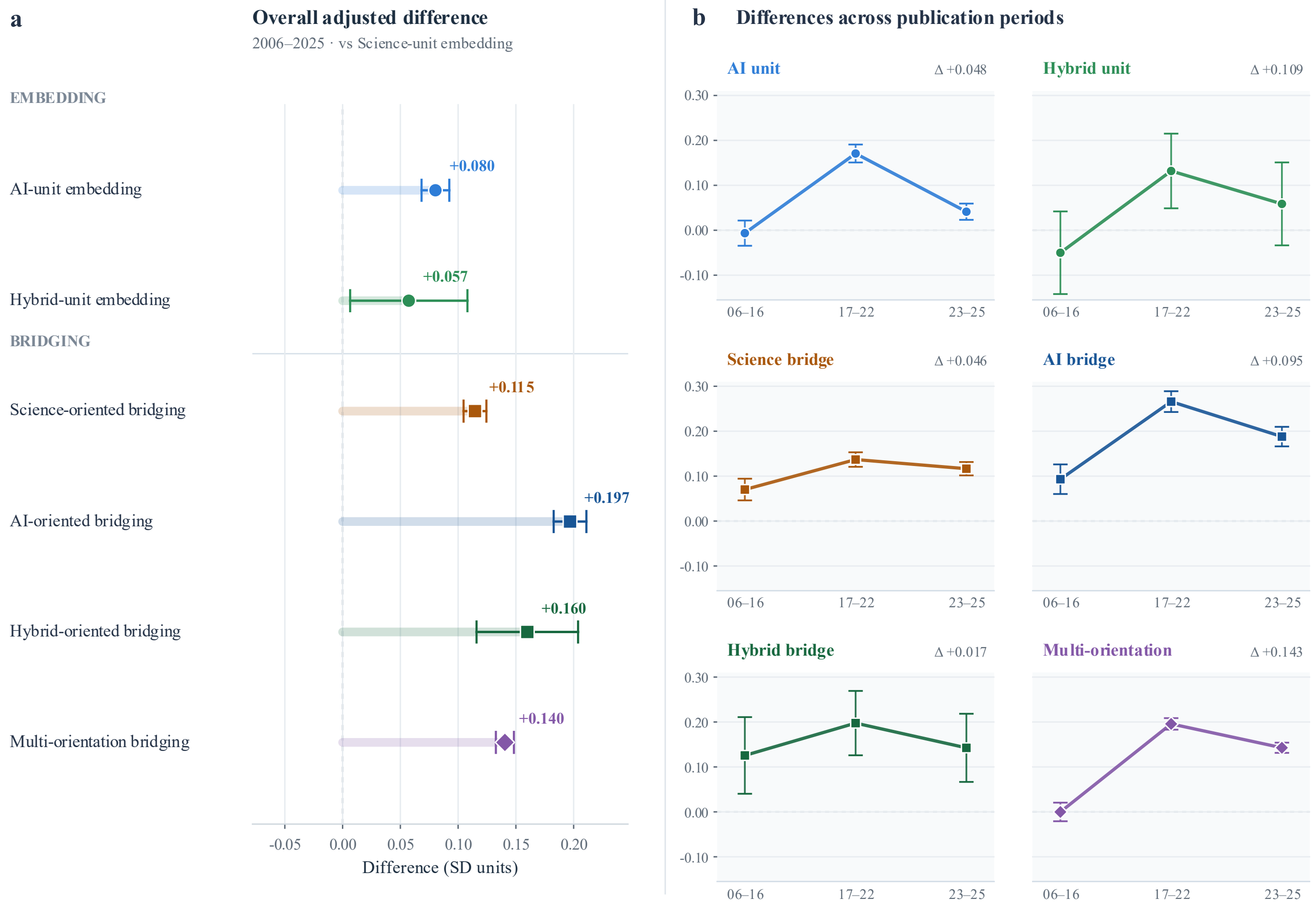

Figure 5. **Organizational configuration and knowledge-recombination breadth.** a, Overall adjusted differences. b, Differences across publication periods.

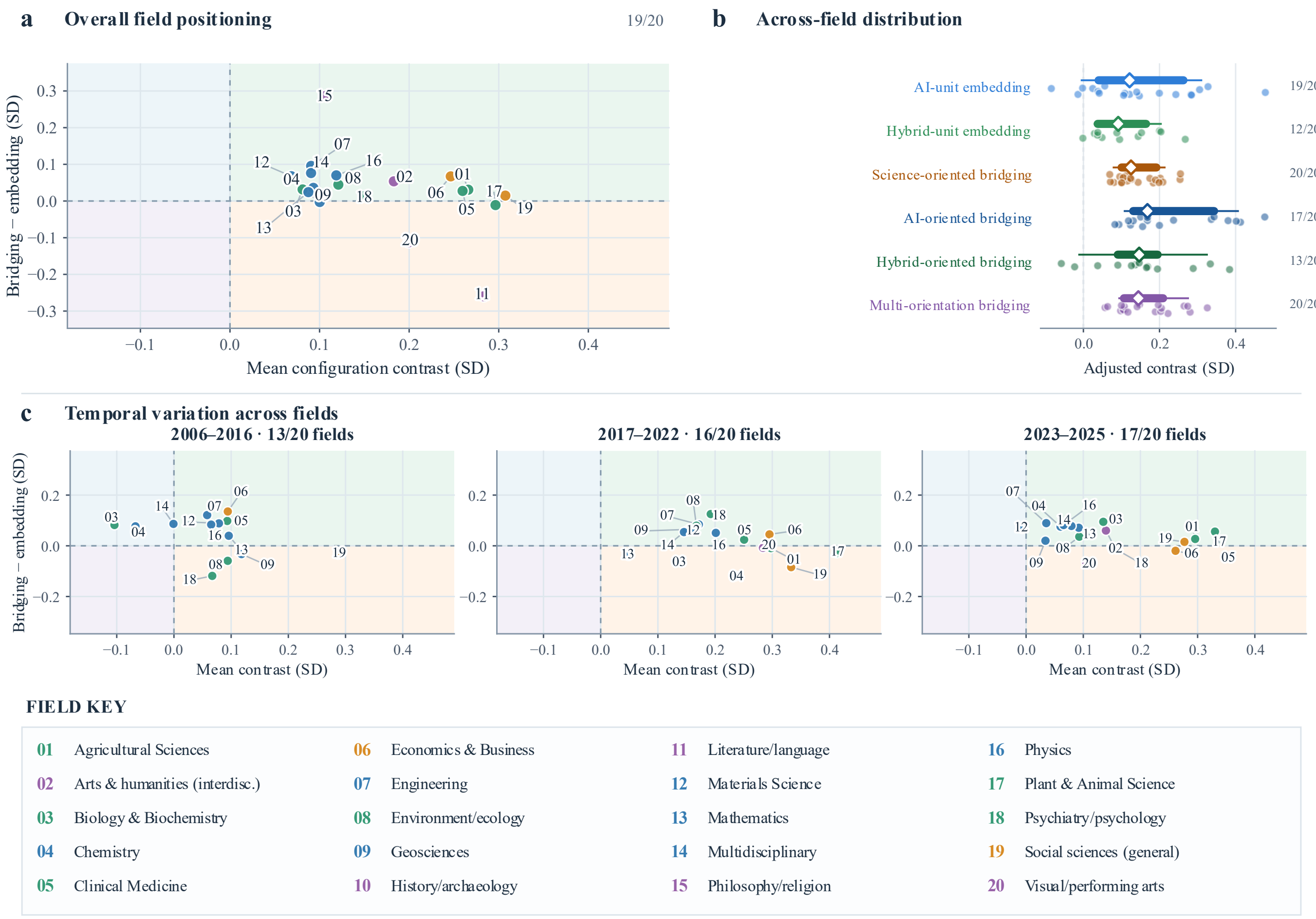

Figure 6. **Field heterogeneity in knowledge-recombination breadth.** a, Overall field positioning. b, Across-field distributions. c, Temporal variation across fields.

### 4.3 Research-impact associations vary across organizational configurations and outcomes

Research impact varies across organizational configurations and outcomes (Figure 7). Science-oriented and Multi-orientation bridging generally exhibit more favorable Q1 publication and citation-impact profiles than Science-unit embedding, particularly from 2017 onward. In 2023–2025, their Q1 publication shares reached 60.2% and 60.5%, respectively, compared with 55.0% for Science-unit embedding. Both configurations also showed higher top-decile citation shares and normalized citation scores in this period, while AI-unit embedding generally exhibited less favorable performance profiles.

These differences largely persist after adjustment for field, publication year, and document type (Supplementary Fig. C4). Relative to Science-unit embedding, Science-oriented and Multi-orientation bridging show adjusted Q1 publication differences of +5.15 and +5.03 percentage points, respectively. Their corresponding differences in top-decile citation probability are +3.36 and +3.21 percentage points, while normalized citation scores are higher by 0.198 and 0.168. AI-unit embedding, by contrast, shows negative differences across all three outcomes. Other bridging configurations also exhibit positive associations, but less consistently across measures. The results therefore suggest that research-performance advantages are associated with particular forms of organizational bridging rather than organizational heterogeneity alone.

Elite-4 publication follows a different pattern. Science-oriented and Multi-orientation bridging are overrepresented among Elite-4 articles in the descriptive profiles (Supplementary Figs. C11–C12), but neither shows a statistically distinguishable adjusted advantage over Science-unit embedding after multiple-testing correction. Several AI- and Hybrid-oriented configurations have lower adjusted Elite-4 probabilities. Thus, the configurations associated with stronger Q1 placement and citation impact do not necessarily have an advantage in publication in these four highly selective journals.

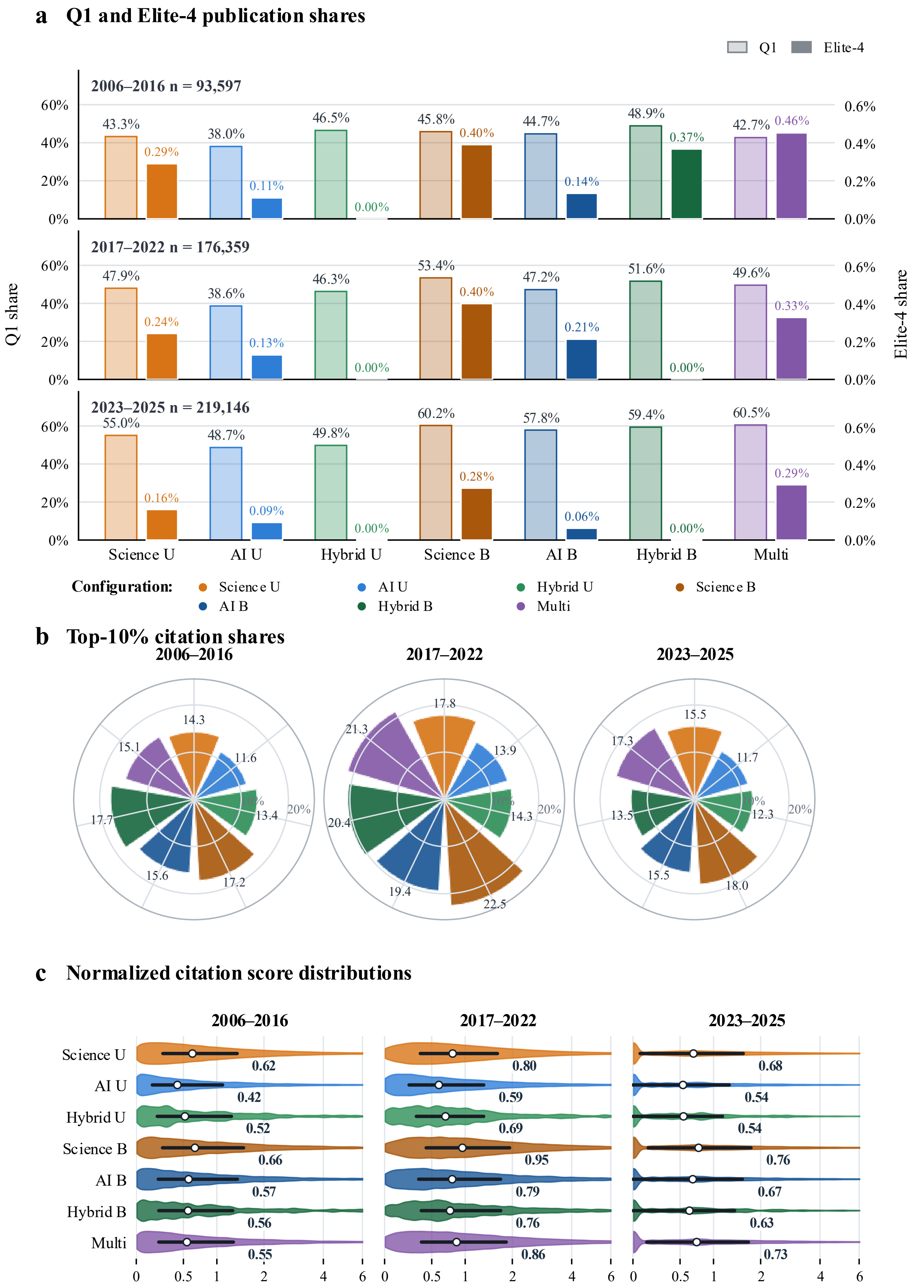

Figure 7. Research-impact profiles across organizational configurations. a, Q1 and Elite-4 publication shares. b, Top-decile citation shares. c, Normalized citation score distributions.

The broader citation-impact pattern extends across disciplinary and national settings, although its magnitude varies. Science-oriented and Multi-orientation bridging show positive adjusted differences in top-decile citation impact and NCS across all ten leading contributing countries, with nominal 95% confidence intervals above zero (Supplementary Figs. C8 and C10). Field-level estimates are more heterogeneous, particularly for AI-unit embedding and less prevalent Hybrid configurations (Supplementary Figs. C7 and C9). Citation-related differences are generally strongest in 2017–2022 and smaller in 2023–2025, when citation accumulation windows are shorter.

### 4.4 Research-impact differences persist after accounting for problem novelty and recombination breadth

The performance differences associated with organizational configuration remain largely intact after accounting for scientific problem novelty and knowledge-recombination breadth (Figure 8). In particular, the positive Q1 publication and citation-impact contrasts associated with Science-oriented and Multi-orientation bridging persist after adjustment for both measures. The adjusted estimates differ only modestly from their baseline values, suggesting that the two knowledge-creation measures account for relatively little of the observed configuration-related differences in research impact.

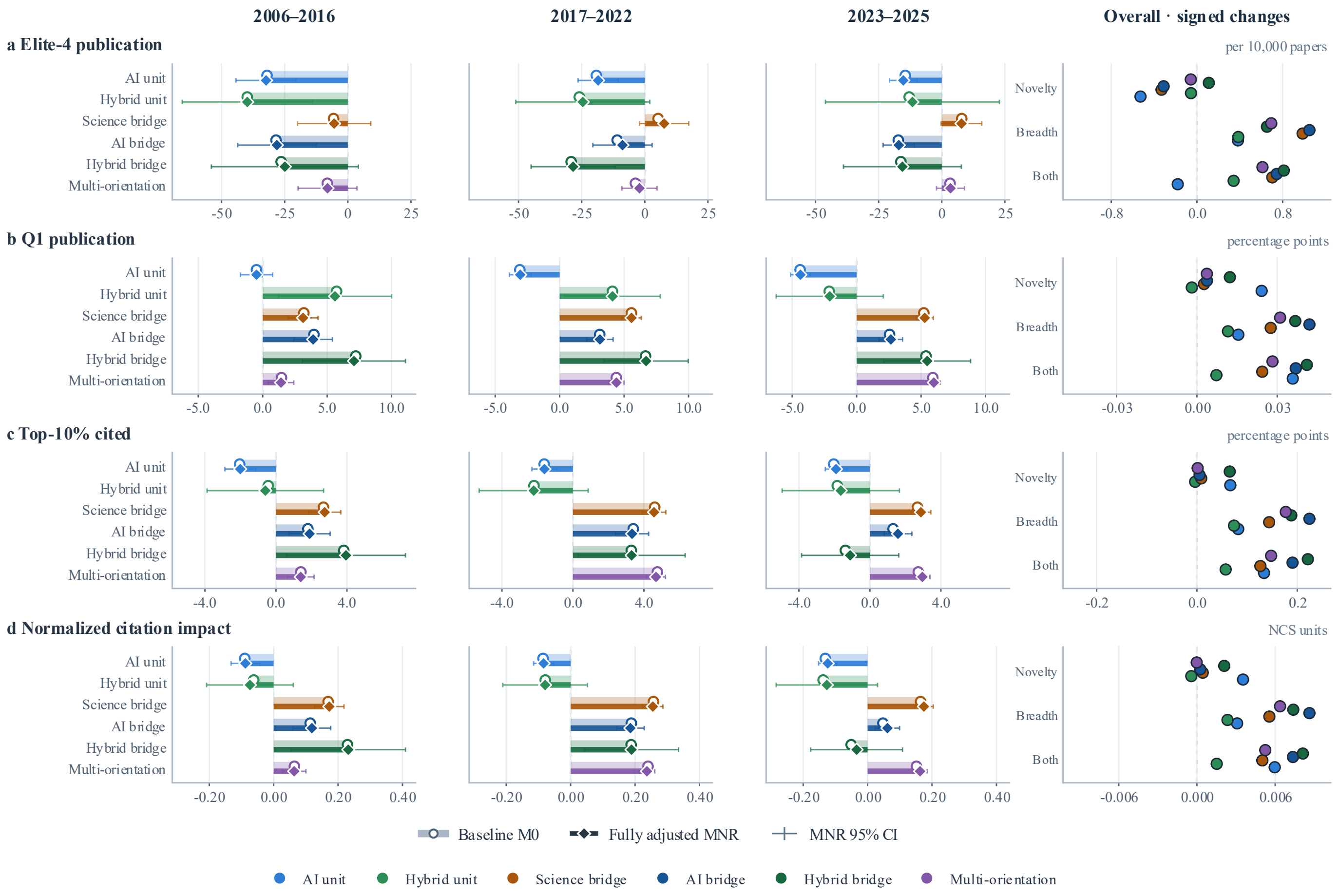

Figure 8. Research-impact differences after adjustment for novelty and recombination breadth. a, Elite-4 publication. b, Q1 journal placement. c, Top-decile citation impact. d, Normalized citation impact.

Adjustment for recombination breadth generally produces larger changes in the estimated contrasts than adjustment for problem novelty, but these changes do not consistently reduce the differences. Several top-decile citation and normalized citation impact contrasts move farther from zero after recombination breadth is included. This is particularly notable given the stronger association between organizational bridging and recombination breadth identified in Section 4.2. Differences in the breadth of knowledge mobilized by organizational configurations do not translate into a corresponding attenuation of their research-performance differences.

The period-specific analyses reinforce this finding (Supplementary Figs. C13–C14). Adjustment for recombination breadth produces larger absolute changes than adjustment for problem novelty in 48 of 72 contrasts. Under joint adjustment, 41 contrasts move closer to zero and 31 move farther away. All 72 retain their original directions, and 71 retain their Holm-adjusted significance status. Although shifts in Elite-4 estimates are less stable because of the rarity of the outcome, the broader results indicate that configuration-related differences in research impact are not substantially accounted for by the two knowledge-creation measures.

## 5. Discussion

This study examines AI4Science not simply as the adoption of computational methods, but as an organizational problem concerning where computational and domain-scientific work is situated and how differently oriented units are connected within scientific production. The results reveal a differentiated pattern rather than a single organizational hierarchy. Organizational configurations differ relatively little in the novelty of the problems pursued, but much more clearly in the breadth of prior knowledge they mobilize. Research performance also varies across configurations, yet the configurations exhibiting the broadest recombination are not necessarily those showing the strongest performance profiles. AI-oriented bridging, for example, exhibits the largest pooled recombination difference, whereas Science-oriented and Multi-orientation bridging show the most consistent positive associations with JCI Q1 publication and citation impact. The divergence between knowledge-recombination breadth and research impact is central to understanding the differentiated performance profiles of AI4Science teams.

### 5.1 Organizational configuration, knowledge creation, and research impact

An organizational perspective on AI4Science reveals differences that cannot be understood from AI adoption alone. The most pronounced distinction among organizational configurations concerns the breadth of knowledge mobilized in research, rather than the novelty of the scientific problems pursued. Cross-unit bridging is associated with broader knowledge recombination, but not with a consistent advantage in problem novelty. One possible explanation is that scientific problems remain rooted in disciplinary agendas, established experimental systems, and prevailing research priorities, while connections between differently oriented units expand the knowledge and methods available to address them. This interpretation is consistent with accounts of path-dependent scientific search (Rzhetsky et al., 2015; Zeng et al., 2019), although the processes through which teams formulate their research problems remain beyond the present analysis.

A second distinction concerns the relationship between knowledge-recombination breadth and research impact. AI-oriented bridging exhibits the broadest recombination, yet Science-oriented and Multi-orientation bridging show more consistently favorable Q1 publication and citation-impact profiles. Moreover, these performance differences remain largely intact after accounting for problem novelty and recombination breadth. Access to heterogeneous knowledge is therefore not equivalent to its productive integration. Crossing knowledge boundaries may expand scientific opportunities while introducing demands for translation, coordination, and validation (Carlile, 2004; Leahey et al., 2017). Whether and how teams meet these demands may matter for research outcomes, but the present measures do not directly capture these processes.

These findings also caution against treating organizational configurations as a hierarchy of research effectiveness. The arrangements associated with broader recombination are not necessarily those associated with stronger performance, and even favorable Q1 and citation-impact profiles do not consistently extend to Elite-4 publication. The contribution of an organizational perspective is thus to distinguish how computational and domain-scientific work is connected from what those connections ultimately produce.

### 5.2 Implications for science policy and research management

The findings suggest that AI4Science policy should move beyond expanding AI capabilities to consider how those capabilities are organized within scientific research. Cross-unit bridging may provide access to more diverse knowledge, but increasing organizational diversity alone cannot be assumed to generate greater problem novelty or stronger research impact. The appropriate organizational arrangement depends on the scientific objectives and the forms of collaboration required to pursue them.

For AI-science research centers, interdisciplinary funding programs, and shared research infrastructures, this calls for greater attention to the organization of collaboration itself. Alongside investments in computational resources and technical expertise, research programs could examine how computational and domain-scientific participants formulate problems, coordinate methodological decisions, and validate scientific findings. These dimensions may help explain why configurations with similarly broad knowledge inputs exhibit different performance profiles.

Research evaluation should therefore distinguish knowledge creation from research impact, rather than treating problem novelty, recombination breadth, journal placement, and citation impact as interchangeable measures of research performance. The present associations do not justify prescribing a particular team structure. They instead provide a basis for developing and evaluating organizational arrangements suited to different research objectives, with attention to disciplinary context and the practical demands of knowledge integration.

### 5.3 Limitations and future research

Several limitations qualify these interpretations. Organizational configurations are inferred from affiliation-unit orientations rather than direct measures of researchers' expertise, organizational capabilities, or collaborative processes. First- and corresponding-author affiliations provide observable indicators of organizational positioning, but they do not necessarily reflect intellectual leadership or research coordination. Authorship conventions vary across disciplines and countries, and WoS metadata do not consistently identify co-first authors, so our classification recognizes only the first-listed author in this role. Future research combining affiliation records with contribution statements, researcher histories, and project-level evidence could better distinguish organizational structure from the actual distribution of expertise and responsibilities.

The knowledge-production measures also capture particular aspects of scientific work rather than scientific value or successful integration. Problem novelty depends on language-model-assisted task extraction and semantic representations, while recombination breadth measures the dispersion of cited knowledge, not how effectively that knowledge is combined. These measures, as well as the broad-AI corpus screening procedure, remain subject to classification and measurement error. The use of JCI quartiles from the 2024 data year may not reflect journals' standing at the time of publication, while recent citation outcomes have shorter accumulation windows. More direct assessments of problem formulation, knowledge integration, and scientific contribution would strengthen the interpretation of these relationships.

Finally, the observational design cannot establish that organizational configuration causes differences in knowledge creation or research impact. Institutional prestige, researcher selection, funding, and project characteristics may influence both organizational arrangements and their observed outcomes. Field and country comparisons also describe heterogeneous associations rather than causal differences between contexts. Future studies could combine project-level organizational data with longitudinal or quasi-experimental designs, using the establishment of AI-science institutes, changes in funding arrangements, or researcher mobility to examine how organizational change affects scientific work.

## 6. Conclusion

AI4Science involves organizational choices as well as technological capabilities. Across seven organizational configurations, the clearest differences lie in the breadth of prior knowledge mobilized rather than the novelty of the scientific problems pursued. Cross-unit bridging is associated with broader knowledge recombination, whereas its associations with research impact vary across organizational configurations. Moreover, differences in problem novelty and recombination breadth do not fully account for the observed performance patterns. For science policy, these findings highlight the importance of understanding how different organizational configurations relate to specific scientific objectives and the conditions under which their potential can be realized, rather than seeking a universally superior organizational model.

## CRediT authorship contribution statement

Yun Fu: Writing - review & editing, Writing - original draft, Visualization, Validation, Project administration, Methodology, Investigation, Formal analysis, Data curation, Conceptualization. Jielan Ding: Writing - review & editing, Project administration, Methodology, Investigation, Conceptualization.

## Declaration of Competing Interest

The authors declare that they have no known competing financial interests or personal relationships that could have appeared to influence the work reported in this paper.

## Funding

This research received financial support from the National Social Science Fund of China (No. 23BTQ103).

## Supplementary materials

The supplementary materials comprise five files: Supplementary_Appendix.docx, containing additional methodological details, validation, robustness analyses, and supplementary results; Supplementary_Note_A1_Research_Task_Extraction_Prompts.md, containing the research-task extraction prompts; Supplementary_Data_1_AI_terms.csv, listing the 2,551 AI search expressions; Supplementary_Data_2_research_task_100_review.csv, providing article-level assessments of research-task extraction; and Supplementary_Data_3_affiliation_300_review.csv, providing unit-level assessments of affiliation-orientation classification.

## Declaration of generative AI and AI-assisted technologies in the manuscript preparation process

During the preparation of this work, the authors used ChatGPT (OpenAI) to assist with manuscript preparation, including language editing and grammatical checks. The authors reviewed and edited the output as needed and take full responsibility for the content of the published article.